\magnification=1200
NON-EQUILIBRIUM STATISTICAL MECHANICS OF TURBULENCE.
\bigskip\noindent
\bigskip\bigskip\bigskip\bigskip
\centerline{by David Ruelle\footnote{$\dagger$}{Math. Dept., Rutgers University, and 
IHES, 91440 Bures sur Yvette, France. email: ruelle@ihes.fr.}.}
\bigskip\bigskip\bigskip\bigskip\noindent
	{\leftskip=1cm\rightskip=1cm\sl {\rm Abstract:}  The macroscopic study of hydrodynamic turbulence is equivalent, at an abstract level, to the microscopic study of a heat flow for a suitable mechanical system [12].  Turbulent fluctuations (intermittency) then correspond to thermal fluctuations, and this allows to estimate the exponents $\tau_p$ and $\zeta_p$ associated with moments of dissipation fluctuations and velocity fluctuations.  This approach, initiated in an earlier note [12], is pursued here more carefully.  In particular we derive probability distributions at finite Reynolds number for the dissipation and velocity fluctuations, and the latter permit an interpretation of numerical experiments [13].  Specifically, if $p(z)dz$ is the probability distribution of the radial velocity gradient we can explain why, when the Reynolds number ${\cal R}$ increases, $\ln p(z)$passes from a concave to a linear then to a convex profile for large $z$ as observed in [13].  We show that the central limit theorem applies to the dissipation and velocity distribution functions, so that a logical relation with the lognormal theory of Kolmogorov [10] and Obukhov is established.  We find however that the lognormal behavior of the distribution functions fails at large value of the argument, so that a lognormal theory cannot correctly predict the exponents $\tau_p$ and $\zeta_p$\par}
\vfill\eject
\noindent
{\bf 1. Introduction.}
\medskip
	In the present paper we study a measure $\varpi$ which describes the velocities in a turbulent 3-D fluid at different levels of a scale of spatial lengths, assuming vanishing average velocities.  The definition of $\varpi$ combines basic ideas of statistical mechanics, the scaling laws of inviscid hydrodynamics, and decorrelation requirements which can hold only approximately.  In particular we assume that the scaling factor $\kappa$ of our  scale of spatial lengths can be chosen such that successive levels of the scale are dynamically decorrelated in a natural manner.  For certain questions we supplement the measure $\varpi$ with a dissipation cutoff involving the viscosity $\nu$.  Since statistical mechanics is involved, $\varpi$ automatically leads to fluctuations of the local velocity (and also of the local energy dissipation, and the local velocity gradients).  Such fluctuations deviate from the homogeneous and isotropic model of turbulence, and have received the name of {\it intermittency}.  One may hope that intermittent fluctuations have a {\it universal} distribution (i.e., a distribution independent of the geometry of the specific turbulent system considered).  In particular, recent numerical experiments [13] support the idea that universality does not require highly developed turbulence, and holds also at relatively small Reynolds number.  The probability measure $\varpi$ discussed in this paper is universal by definition, and the problem we address here is to what extent $\varpi$ fits the data provided by (lab and computer) experiments.  A numerical approach to this problem is certainly desirable, but we shall here proceed analytically, taking advantage of the very explicit form of $\varpi$ (see equation (3) below).  We shall make many approximations, based in particular on the fact that the scaling constant $\kappa$ is ``large'' (between 20 and 25), and we shall make liberal use of the notation $\approx$ (approximately equal).  The spirit of our approach is thus to start from something very robust: the basic ideas of statistical mechanics (usually ignored in turbulence theory) and see to what extent a connection with experiments is achieved in spite of the crude approximations that we shall make.  The results appear rather encouraging: this reflects probably the fact that the quantities of interest are {\it rates} associated with exponentially behaving quantities (such rates are typically rather stable under approximations).
\medskip
	We have thus evidence that our approach to turbulence, based on ideas of statistical mechanics, is basically correct.  This leads in particular to some useful conclusions concerning the lognormal turbulence theory of Kolmogorov and Obukhov.  From $\varpi$ one obtains convolution product expressions (4) and (14) below, to which one can apply the central limit theorem: this gives approximately lognormal distributions for the local dissipation and the radial velocity increment.  But the convolution products just referred to involve functions $\alpha$, $\phi$, $\psi$ which are explicitly known, and one can check that the asymptotic exponents $\tau_p,\zeta_p$ associated with the local energy dissipation and velocity increment distributions are not those predicted by the lognormal theory.  One could say that the lognormal theory contains an element of truth but has limited applicability.
\medskip
	We review now the theory of hydrodynamic turbulence as proposed here, following an earlier paper [12].  In this theory,  the fluid system is represented as an interacting union of subsystems $(n,i)$ with finite degrees of freedom, in such a way that the turbulent energy cascade corresponds to a heat flow through the collection of subsystems.  Specifically, the fluid system is enclosed in a cube $C_0$, and for each positive integer $n$, $C_0$ is cut in $\kappa^{3n}$ subcubes $C_{ni}$ with edge size $\ell_n=\ell_0\kappa^{-n}$ for some choice of $\kappa$.  A wavelet representation of divergence-free velocity fields provides a description of the inviscid Hamiltonian for the fluid in $C_0$ as an interacting collection of subsystems $(n,i)$ roughly localized in the cubes $C_{ni}$.  For the study of the turbulent energy cascade, one assumes that energy is input at large spatial wavelengths and dissipated at small spatial wavelengths.  In the description in terms of the systems $(n,i)$ the turbulent cascade corresponds to a heat flow from $n=0$ to $n=N$ for large $N$ (we only discuss the Hamiltonian {\it inertial range}).  This means roughly that the temperature\footnote{$^*$}{A different concept of temperature in the study of turbulence has been introduced by B. Castaing [4], as kindly pointed out by F. Bouchet, but Castaing's temperature is conserved along the cascade, contrary to the situation considered here.} at $n=0$ is kept larger than the temperature at $n=N$ (in fact the heat flow from $n=0$ is fixed rather than the temperature at $n=N$).  Translating the study of the turbulent cascade into a problem of heat flow replaces the question of velocity fluctuations (i.e., intermittency) by a question of fluctuations in non-equilibrium statistical mechanics  (in principle a very hard question).  This was handled in [12] by arguing that the subsystems $(n,i)$ are in approximate thermal equilibrium so that one can use the Boltzmann distribution to study fluctuations.
\medskip
	Let us proceed with a more precise discussion.  We take our fluid to be contained in a cube $C_0$ of side $\ell_0$ and let the velocity field ${\bf v}$ satisfy $\int{\bf v}=0$, ${\rm div}\,{\bf v}=0$.  (A description in terms of the vorticity might be preferred, but would not change the present discussion).  We divide $C_0$ into cubes $C_{ni}$ of side $\ell_n=\ell_0\kappa^{-n}$ (with $i=1,\ldots,\kappa^{3n}$) and denote by $\phi_{ni}$ the homothety mapping $C_0$ to $C_{ni}$.  Choosing $2(\kappa^3-1)$ real vector fields ${\bf U}_\alpha$ on ${\bf R}^3$ with $\int{\bf U}_\alpha=0$, ${\rm div}\,{\bf U}_\alpha=0$, we assume that ${\bf v}$ has a unique wavelet decomposition into components (roughly) localized in the cubes $C_{ni}$:
$$	{\bf v}=\sum_{n=0}^\infty\sum_{i=1}^{\kappa^{3n}}\sum_{\alpha=1}^{2(\kappa^3-1)}
	c_{ni\alpha}{\bf U}_\alpha\circ\phi_{ni}^{-1}   $$
with $c_{ni\alpha}\in{\bf R}$.  We write ${\bf v}_{ni}=\sum_\alpha c_{ni\alpha}{\bf U}_\alpha\circ\phi_{ni}^{-1}$.  Notice that the values of ${\bf v}_{ni}$ at the centers of the $\kappa^3$ cubes $C_{(n+1)j}\subset C_{ni}$ are not independent (there are only $\approx(2/3)(\kappa^3-1)$ independent values).
\medskip
	Note that according to Kolmogorov theory $|{\bf v}_{ki}|\sim(\epsilon\ell_k)^{1/3}$, where $\epsilon$ is the mean energy dissipation per unit volume.
\medskip
	Energy conservation is expressed (as in the multifractal approaches, see [5], [3], [11], [6]) by
$$	{|{\bf v}_{ni}|^3\over\ell_n}={|{\bf v}_{(n+1)j}|^3\over\ell_{n+1}}\qquad{\rm or}
	\qquad|{\bf v}_{(n+1)j}|^3={|{\bf v}_{ni}|^3\over\kappa}\eqno{(1)}   $$
This is because the kinetic energy $|{\bf v}|^2/2$ in a given spatial range is weighted by the inverse of the  time $t$ spent in this range, and $t$ scales like $\ell/|{\bf v}|$.  We interpret $(1)$ to mean that, corresponding to a given ${\bf v}_{ni}$, there corresponds a fluctuating ${\bf v}_{(n+1)j}$ such the the average of $V_{n+1}=|{\bf v}_{(n+1)j}|^3$ is $V_n/\kappa=|{\bf v}_{ni}|^3/\kappa$.  It is natural to assume that the distribution of ${\bf v}={\bf v}_{(n+1)j}$ in ${\bf R}^3$ maximizes the entropy, and is thus the Boltzmann distribution
$$	\sim\exp\Big(-{|{\bf v}|^3\over V_n\kappa^{-1}}\Big)d^3{\bf v}   $$
i.e., $V=V_{n+1}$ has the distribution
$$	{1\over V_n\kappa^{-1}}\exp\Big(-{V\over V_n\kappa^{-1}}\Big)dV\eqno{(2)}   $$
This is the choice that was made in [12].  It is a canonical ensemble expression expected to be reasonable\footnote{$^*$}{While the $|{\bf v}_{(n+1)j}|^3/\ell_{n+1}$ fluctuate, their sum over $j$ is fixed ($\sim$ total energy flux in $C_{ni}$) so that very large values of $V$ are forbidden in $(2)$.  This is related to the fact that fixing the average of the ``energies'' $|{\bf v}_{(n+1)j}|^3$ corresponds to a microcanonical ensemble, which is only asymptotically equivalent to the canonical ensemble $(2)$, but the approximation should be good for moderate $V$ and large $\kappa$.} for moderately large $V$ and large $\kappa$.
\medskip
	Note that we consider only interactions between subsystems $(n,i)$, $(n+1,j)$ such that $C_{ni}\supset C_{(n+1)j}$.  This is a strong form of the locality (in $k$ space) usually assumed in turbulence theory.  The lack of interaction between $(n,i)$, $(n,i')$ for $i\ne i'$ means that momentum conservation is not properly taken into account.  In this respect we are following the multifractal approaches [5], [3], [11], [6].  The multifractal approaches are purely ad hoc, but physically motivated modifications have recently been introduced [15], [14].  The approach presented here has physical justification but makes the limiting assumption that the local average velocity of the fluid vanishes.  For the study of concrete problems it will be necessary to make more general assumptions, taking into account the geometry of the situation considered.  This can in principle be done, and will hopefully  lead to more concrete studies of turbulence based on the physical ideas of nonequilibrium statistical mechanics.
\medskip
	In [12] we applied the ideas sketched above to the study of the exponents $\zeta_p$ associated with the moments $\langle|\Delta_{\bf r}{\bf v}|^p\rangle$ of velocity increments.  In the present paper we study, in Section 2, the fluctuations of the scale dependent energy dissipation $\epsilon_\ell$ and the corresponding exponents $\tau_p$.  In Section 3 we study the probability distribution of the the radial velocity increments $\Delta_rv$.  Note that the exponents $\tau_p,\zeta_p$ are large Reynolds number limits, while the distributions of the fluctuations of the energy dissipation and the radial velocity increments obtained in Sections 2 and 3 may be compared with experiments at finite Reynolds number.  We discuss small radial velocity increments in Section 4, and large velocity increments in Section 5, obtaining an explanation of profiles obtained in [13].  The relation of the present paper with the lognormal theory of Kolmogorov [10] and Obukhov, is discussed in Section 6.
\vfill\eject
\medskip\noindent
{\bf Acknowledgments.}
\medskip
	I am indebted to Christian Beck and Victor Yakhot for useful interaction during the preparation of this paper, and to Giovanni Gallavotti for earlier discussions; VY communicated reference [13] which is essential here.  This work was initiated while I visited the Isaac Newton Institute in Cambridge, UK, at the end of 2013.
\medskip\noindent
{\bf 2. Fluctuations of the energy dissipation.}
\medskip
	We follow (1) and (2), and write $V_{ni}=|{\bf v}_{ni}|^3$.  Then the normalized distribution of $V=V_{(n+1)j}$ is given by (2) (we omit henceforth the index $i$).  We find thus that, corresponding with the chain $C_0\supset C_1\supset\cdots\supset C_n$, with fixed $V_0$, there is a probability distribution
$$	{\kappa\over V_0}e^{-\kappa V_1/V_0}{\kappa\over V_1}e^{-\kappa V_2/V_1}\cdots
	{\kappa\over V_{n-1}}e^{-\kappa V_n/V_{n-1}}dV_1\cdots dV_n\eqno{(3)}   $$
and this extends to a probability measure $\varpi$ on the space of infinite sequences $(V_n)_1^\infty$.
In agreement with (1) we define the average energy dissipation $\epsilon_{\ell_n}$ at size $\ell_n$ to be $V_n/\ell_n$ (this is an approximation of the definition in [10]).  Writing $w_k=\kappa V_k/V_{k-1}$ (so that $\kappa^k V_k=V_0w_1\cdots w_k$) we have thus
$$	\langle\epsilon_{\ell_n}^p\rangle=\int{\kappa\,dV_1\over V_0}e^{-\kappa V_1/V_0}
	\int{\kappa\,dV_2\over V_1}e^{-\kappa V_2/V_1}\cdots
	\int{\kappa\,dV_n\over V_{n-1}}e^{-\kappa V_n/V_{n-1}}\Big({V_n\over\ell_n}\Big)^p   $$
$$	=\Big({V_0\over\ell_0}\Big)^p\int d w_1\,e^{-w_1}\int dw_2\,e^{-w_2}\cdots\int dw_n\,e^{-w_n}
	(w_1w_2\cdots w_n)^p   $$
$$	=\Big({V_0\over\ell_0}\Big)^p\Big(\int w^pe^{-w}\,dw\Big)^n
	=\Big({V_0\over\ell_0}\Big)^p\Big(\Gamma(1+p)\Big)^n   $$
so that for large $n$
$$	{\ln\langle\epsilon_{\ell_n}^p\rangle\over\ln\ell_n}
	={p\ln(V_0/\ell_0)+n\ln\Gamma(1+p)\over\ln\ell_0-n\ln\kappa}\approx-{\ln\Gamma(1+p)\over\ln\kappa}   $$
and we obtain $\langle\epsilon_{\ell_n}^p\rangle\approx\ell_n^{\tau_p}$ with
$$	\tau_p=-{\ln\Gamma(1+p)\over\ln\kappa}\qquad\hbox{so that}\qquad\zeta_p={p\over3}+\tau_{p/3}   $$
where we have used the value of $\zeta_p$ obtained in [12].
\medskip
	Note by the way that the validity of (3) is limited by the dissipation due to the viscosity $\nu$, i.e., we must have $V_n^{1/3}\ell_n>\nu$; this will be used later.
\medskip
	Let $D_n(x)\,dx$ be the distribution, given $V_0$, of $x=\epsilon_{\ell_n}/\epsilon_{\ell_0}$ on ${\bf R}_+$.  According to (3) we have
$$	D_n(x)=\int{\kappa dV_1\over V_0}e^{-\kappa V_1/V_0}\int{\kappa V_2\over V_1}e^{-\kappa V_2/V_1}
\cdots\int{\kappa dV_n\over V_{n-1}}e^{-\kappa V_n/V_{n-1}}\,\delta(x-{V_n\over\ell_n}\cdot{\ell_0\over V_0})   $$
$$	=\int dw_1\,e^{-w_1}\int dw_2\,e^{-w_2}\cdots\int dw_n\,e^{-w_n}\,\delta(x-w_1\cdots w_n)   $$
so that, writing $\alpha(t)=\exp(t-e^t)$, we have
$$	e^tD_n(e^t)=e^t\int\cdots\int\delta(e^t-e^{t_1+\ldots+t_n})\prod_{k=1}^n\Big(\exp(t_k-e^{t_k})\,dt_k\Big)   $$
$$	=\int\cdots\int\delta(t_1+\ldots+t_n-t)\prod_{k=1}^n(\alpha(t_k)\,dt_k)=\alpha^{*n}(t)\eqno{(4)}   $$
where $*$ denotes the convolution product.  The central limit theorem implies thus that $e^tD_n(e^t)\,dt$ is asymptotically Gaussian for large $n$, i.e., that $D_n(x)\,dx$ is asymptotically a lognormal distribution.  Therefore the distribution of $\epsilon_{\ell_n}$ (given $V_0$) is asymptotically lognormal.
\medskip
	A word of caution is needed here: when $t\to-\infty$ we have $\alpha(t)\approx\eta_1(-t)=\eta_1(|t|)$, where we have written $\eta_m(t)=\theta(t)e^{-mt}$ and we have $\eta_m^{*n}(t)=\theta(t)t^{n-1}e^{-mt}/(n-1)!$  Therefore, when $t\to-\infty$ we have $e^tD_n(e^t)\approx\eta_1^{*n}(|t|)$ and, for small $x$, $D_n(x)$ diverges like $|t|^{n-1}/(n-1)!=|\ln x|^{n-1}/(n-1)!$  This divergence is however cut off by the condition $V_n^{1/3}\ell_n>\nu$, or $\epsilon_n=V_n/\ell_n=(V_n^{1/3}\ell_n)^3/\ell_n^4>\nu^3/\ell_n^4=\kappa^{4n}\nu^3/\ell_0^4$, i.e.,
$$	V_n^{1/3}\ell_n>\nu\qquad\Leftrightarrow\qquad{\epsilon_n\over\epsilon_0}
	=x>{\kappa^{4n}\nu^3\over V_0\ell_0^3}=\kappa^{4n}{\cal R}^{-3}\eqno{(5)}   $$
where ${\cal R}=V_0^{1/3}\ell_0/\nu$ is the Reynolds number.
\medskip
	We have
$$	\alpha^{*n}(t)=\int dt_1\cdots\int dt_n\,\delta(t_1+\ldots+t_n-t)\prod_{k=1}^n\exp(t_k-e^{t_k})   $$
$$	=e^t\int dt_1\cdots\int dt_n\,\delta(t_1+\ldots+t_n-t)\exp(-\sum_{k=1}^ne^{t_k})   $$
$$	=e^t\int dt'_1\cdots\int dt'_n\,\delta(t'_1+\ldots+t'_n)\exp(-e^{t/n}\sum_{k=1}^ne^{t'_k})   $$
A quadratic approximation of $\sum e^{t'_k}$ gives
$$	\sum_{k=1}^ne^{t'_k}\approx\sum_{k=1}^n(1+t'_k+{1\over2}{t'_k}^2)
=	n+{1\over2}\sum_{k=1}^n{t'_k}^2=n+{1\over2}[\sum_{k=1}^{n-1}{t'_k}^2+(\sum_{k=1}^{n-1}t'_k)^2]
=	n+{1\over2}t'^TAt'   $$
where $t'\in{\bf C}^{n-1}$ has components $t'_1,\ldots,t'_{n-1}$ and $A$ is a $(n-1)\times(n-1)$ matrix with $A_{k\ell}=2$ if $k=\ell$, $=1$ otherwise.  One finds $\det A=n$, and we have thus
$$	\alpha^{*n}(t)\approx e^t\int dt'_1\cdots\int dt'_{n-1}\exp[-e^{t/n}(n+{1\over2}t'^TAt')]   $$
$$	=e^t\exp(-ne^{t/n})\Big({(2\pi)^{n-1}\over ne^{t(n-1)/n}}\Big)^{1/2}
	=\Big({(2\pi)^{n-1}\over n}\Big)^{1/2}e^{t(1/n+1)/2}\exp(-ne^{t/n})   $$
$$	D_n(x)\approx\Big({(2\pi)^{n-1}\over n}\Big)^{1/2}x^{(1/n-1)/2}\exp(-nx^{1/n})   $$
and for large $X$
$$	\int_X^\infty D_n(x)\,dx\approx\Big({(2\pi)^{n-1}\over n}\Big)^{1/2}(X^{(1-1/n)/2}+\ldots)\exp(-nX^{1/n})   $$
Therefore, writing $S_j=\{(V_n)_1^\infty:V_j^{1/3}\ell_j>\nu\}$ we have, using (5),
$$   \ln\varpi(S_n)=\ln\int_{\kappa^{4n}{\cal R}^{-3}}^\infty D_n(x)\,dx\approx-n\kappa^4{\cal R}^{-3/n}\eqno{(6)}   $$
so that $\varpi(S_n)$ decreases exponentially with $n$.
\medskip
	We shall use later the results
$$	\int t\alpha(t)\,dt=\int_0^\infty(\ln x)e^{-x}\,dx=-\gamma\quad,\quad
	\int t^2\alpha(t)\,dt=\int_0^\infty(\ln x)^2e^{-x}\,dx=\gamma^2+{\pi^2\over6}   $$
where $\gamma$ is Euler's constant, so that ${\rm Var}(\alpha)=\pi^2/6$.
\medskip\noindent
{\bf 3. Fluctuations of velocity increments.}
\medskip
	This Section contains the main technical machinery of the present paper.  Contrary to the study of the exponents $\tau_p,\zeta_p$ we shall not take the limit of infinite Reynolds number ($n\to\infty$).
\medskip
	Here we study the distribution of $\Delta_{\bf r}{\bf v}={\bf v}({\bf x}+{\bf r})-{\bf v}({\bf x})$, assuming that ${\bf x}$ and ${\bf x}+{\bf r}$ are in the same cube $C_{k,i(k)}$ for $k\le n$, and in different cubes for $k>n$.  Then $\Delta_{\bf r}{\bf v}=\sum_k\Delta_k{\bf v}$ where
$$	\Delta_k{\bf v}={\bf v}_{k,i(k)}({\bf x}+{\bf r})-{\bf v}_{k,i(k)}({\bf x})\qquad\hbox{for $k\le n$}   $$
$$	\Delta_k{\bf v}={\bf v}_{k,i'(k)}({\bf x}+{\bf r})-{\bf v}_{k,i(k)}({\bf x})\qquad\hbox{for $k>n$}   $$
We see that $|\Delta_k{\bf v}|$ must tend to $0$ when $|k-n|$ grows.  A simple approximation is thus to replace $\Delta_{\bf r}{\bf v}=\sum_k\Delta_k{\bf v}$ by $\Delta_n{\bf v}$, and $\Delta_n{\bf v}$ by ${\bf v}_n={\bf v}_{n,i(n)}$ as was done in [12] to estimate the exponents $\zeta_p$ of $\langle|{\bf v}_n|^p\rangle\approx\ell_n^{\zeta_p}$ (as earlier we write ${\bf v}_k$ instead of ${\bf v}_{k,i(k)}$).
\medskip
	We shall now study specifically the distribution of the {\it radial increment} $\Delta_rv$ obtained by choosing a coordinate axis in ${\bf R}^3$ (the $x$-axis), taking ${\bf r}$ along the $x$-axis, and letting $r$, $\Delta_rv$ be the components of ${\bf r}$, $\Delta_{\bf r}{\bf v}$ along the $x$-axis.  The ideas will be the same as in [12], presented a bit more carefully, and will give interesting information on the distribution of large values of  $\Delta_rv$.  Our discussion will be approximate, in particular the distribution obtained will be symmetric with respect to the reflection $\Delta_rv\to-\Delta_rv$.
\medskip
	In the continuous limit we can think of the vector field ${\bf U}_\alpha$ used to represent ${\bf v}$ as given by ${\bf U}_\alpha({\bf x})={\bf j}e^{i{\bf k}\cdot{\bf x}}$ with ${\bf j}\cdot{\bf k}=0$.  If ${\bf k}$ is uniformly distributed on the sphere $|{\bf k}|=1$ and ${\bf j}$ on the circle $|{\bf j}|=1$, the component $u$ of ${\bf j}$ along a coordinate axis in ${\bf R}^3$ (the $x$-axis) has a distribution $\sim du/\sqrt{\alpha^2-u^2}$ where $\alpha^2=1-\beta^2$ and $\beta$ is uniformly distributed on $[-1,1]$.  Therefore the $x$-component $u$ of ${\bf U}_\alpha$ has a distribution
$$	\sim du\int_0^{\sqrt{1-u^2}}{d\beta\over\sqrt{1-u^2-\beta^2}}
	=du\int_0^1{d\gamma\over\sqrt{1-\gamma^2}}\sim du\eqno{(7)}   $$
on $[-1,1]$.  We shall use this fact in a moment.
\medskip
	If we fix a velocity field distribution $\mu(d{\bf v}_0)$ in $C_0$ we can, using (3), compute the average of a function $\Phi((|{\bf v}_k|^3)_{k=0}^N)$ by
$$	\langle\Phi\rangle=\int_0^\infty[\int\mu(d{\bf v}_0)\,\delta(|{\bf v}_0|^3-V_0)]F(V_0)   $$
$$	F(V_0)=\Big(\prod_{k=1}^N\int_0^\infty{\kappa dV_k\over V_{k-1}}e^{-\kappa V_k/V_{k-1}}\Big)
	\Phi(V_0,V_1,\ldots,V_N)\eqno{(8)}   $$
\par
	Given $V_1,\ldots,V_n$, the probability distribution of the $x$-components $v_1,\ldots,v_N$ of ${\bf v}_1$, $\ldots,{\bf v}_N$  is, in view of (7), given by
$$	\Phi(v_1,\ldots,v_N;V_1,\ldots,V_N)\,dv_1\cdots dv_N
	=\prod_{k=1}^N\Big({dv_k\over2V_k^{1/3}}\,\chi_{[-V_k^{1/3},V_k^{1/3}]}(v_k)\Big)\eqno{(9)}   $$
where $\chi_A$ denotes the characteristic function of $A$.  This particular choice of $\Phi$ corresponds by (8) to
$$	F(v_1,\ldots,v_N;V_0)=\prod_{k=1}^N\Big(\int_0^\infty{\kappa dV_k\over V_{k-1}}e^{-\kappa V_k/V_{k-1}}	\cdot{1\over2V_k^{1/3}}\cdot\chi_{[-V_k^{1/3},V_k^{1/3}]}(v_k)\Big)\eqno{(10)}  $$
More generally, if $\lambda_1,\ldots,\lambda_N>0$ we may study the probability that the $x$-components of $\lambda_1{\bf v}_1,\ldots,\lambda_N{\bf v}_N$ are $u_1,\ldots,u_N$.  Then $(9)$ and $(10)$ are replaced by
$$	\Phi_\lambda(u_1,\ldots,u_N;V_1,\ldots,V_N)\,du_1\cdots du_N
=\prod_{k=1}^N\Big({du_k/\lambda_k\over2V_k^{1/3}}\,\chi_{[-V_k^{1/3},V_k^{1/3}]}(u_k/\lambda_k)\Big)   $$
$$	F_\lambda(u_1,\ldots,u_N;V_0)
	=\prod_{k=1}^N\Big(\int_0^\infty{\kappa dV_k\over V_{k-1}}e^{-\kappa V_k/V_{k-1}}\cdot
{1\over2\lambda_kV_k^{1/3}}\cdot\chi_{[-\lambda_kV_k^{1/3},\lambda_kV_k^{1/3}]}(u_k)\Big)\eqno{(11)}  $$
\medskip
	We have denoted above by $r$, $v_k$ the $x$-components of ${\bf r}$, ${\bf v}_k$; let also $u_k$ be the $x$-components of $\Delta_k{\bf v}$.  We claim that the probability distribution of the velocity increment $u_k$ is roughly the same as the distribution of a suitable multiple $\lambda_kv_k$ of $v_k$, as follows:
\par
	if $k\le n$:$\qquad\lambda_k\approx\kappa\ell_{n+1}/\ell_k$
\par
	if $k\ge n$:$\qquad\lambda_k\approx1$
\par\noindent
This simply expresses the fact that $r\approx\ell_{n+1}$ and that the correlation length of $v_k$ is $\approx\ell_k/\kappa$.  We may thus apply $(11)$ with $\lambda_k$ as above.
\medskip
	Let $F_\Delta(u;V_0)\,du$ denote the probability distribution of $u=\Delta_rv$ for given $V_0$.  We have thus
$$	F_\Delta(u;V_0)
	=\Big(\prod_{k=1}^N\int_0^\infty{\kappa dV_k\over V_{k-1}}e^{-\kappa V_k/V_{k-1}}\Big)
	\Phi_\Delta(u;V_1,\ldots,V_N)   $$
where, with the notation indicated,
$$	\Phi_\Delta(u;V_1,\ldots,V_N)=\Big(\prod_{k=1}^N\int{du_k
\over{2\lambda_kV_k^{1/3}}}\,\chi_{[-\lambda_kV_k^{1/3},\lambda_kV_k^{1/3}]}(u_k)\Big)\delta(u-\sum u_k)   $$
In view of the values of the $\lambda_k$, a rough estimate of $\Phi_\Delta$ is obtained by taking $\sum u_k=u_n$ so that\footnote{*}{Notice that $\Phi_\Delta$ is a convolution product with respect to the variable $u$: replacing $\sum u_k$ by $u_n$ replaces $N-1$ factors of this convolution product by a Dirac $\delta$.}
$$	\Phi_\Delta(u;V_1,\ldots,V_N)\approx{1\over{2V_n^{1/3}}}\,\chi_{[-V_n^{1/3},V_n^{1/3}]}(u)
	={1\over{2V_n^{1/3}}}\,\chi_{[|u|^3,\infty)}(V_n)   $$
and
$$	F_\Delta(u;V_0)\approx F_\Delta(u)=\Big(\prod_{k=1}^n\int_0^\infty{\kappa dV_k\over V_{k-1}}
	e^{-\kappa V_k/V_{k-1}}\Big){1\over{2V_n^{1/3}}}\,\chi_{[-V_n^{1/3},V_n^{1/3}]}(u)   $$
Writing $W_k=\kappa^k V_k$ and $w_k=W_k/W_{k-1}$ (so that $W_k=V_0w_1\cdots w_k$) we find
$$	F_\Delta(u)=\Big(\prod_{k=1}^n\int_0^\infty{dW_k\over W_{k-1}}e^{-W_k/W_{k-1}}\Big)
	{1\over2(\kappa^{-n}W_n)^{1/3}}\,\chi_{[|u|^3,\infty)}(\kappa^{-n}W_n)   $$
$$	={\kappa^{n/3}\over2}\Big(\prod_{k=1}^{n-1}\int_0^\infty{dW_k\over W_{k-1}}e^{-W_k/W_{k-1}}\Big)
	\int_{\kappa^n|u|^3}^\infty{dW_n\over W_{n-1}}\,e^{-W_n/W_{n-1}}\,{1\over W_n^{1/3}}   $$
$$	={\kappa^{n/3}\over2}\Big(\prod_{k=1}^{n-1}\int_0^\infty dw_k\,e^{-w_k}\Big)
	\int_{\kappa^n|u|^3/V_0w_1\cdots w_{n-1}}dw_n\,e^{-w_n}\,{1\over(V_0w_1\cdots w_n)^{1/3}}   $$
$$	={1\over2}\big({\kappa^n\over V_0}\big)^{1/3}\int\cdots\int_{w_1\cdots w_n>(\kappa^n/V_0)|u|^3}
	\prod_{k=1}^n{dw_k\,e^{-w_k}\over w_k^{1/3}}   $$
\par
	Since $F_\Delta(u)$ is an even function of $u$ it is natural to consider the probability distribution $G_n(y)dy$ of $y=(\kappa^n/V_0)^{1/3}|u|$ corresponding to $F_\Delta(u)\,du$.  We have
$$	G_n(y)=\int\cdots\int_{w_1\cdots w_n>y^3}\prod_{k=1}^n{dw_k\,e^{-w_k}\over w_k^{1/3}}   $$
$$	G_n(e^t)=\int\cdots\int_{t_1+\cdots+t_n>t}\prod_{k=1}^n(3\exp(2t_k-e^{3t_k})\,dt_k)   $$
$$	=\int_{-\infty}^{+\infty}\cdots\int_{-\infty}^{+\infty} dt_1\cdots dt_{n-1}\int_{t-t_1-\cdots-t_{n-1}}^{+\infty}dt_n
	\prod_{k=1}^n(3\exp(2t_k-e^{3t_k}))\eqno{(12)}   $$
\par
	Define
$$	\phi(t)=3\exp(3t-e^{3t})\qquad,\qquad\psi(t)=e^t\int_t^\infty3\exp(2s-e^{3s})\,ds   $$
so that
$$	\phi(t)=-e^t{d\over dt}(e^{-t}\psi(t))\qquad,\qquad\psi(t)=e^t\int_t^\infty e^{-s}\phi(s)\,ds   $$
and $\phi(t)dt,\psi(t)dt$ are probability distributions on ${\bf R}$.  Note that
$$	e^{-t}\psi(t)=\int_{e^t}^\infty3vdv\,e^{-v^3}=\int_{e^{3t}}^\infty{dw\over w^{1/3}}e^{-w}
	=\Gamma({2\over3},e^{3t})   $$
where $\Gamma(a,x)$ is the upper incomplete Gamma function, so that $\psi(t)\approx\Gamma(2/3)e^t$ for $t\to-\infty$, $\approx\exp(-e^{3t})$ for $t\to\infty$.
\medskip
	Using $(12)$ we find the convolution expression:
$$	-e^t{d\over dt}G_n(e^t)=e^t\int\cdots\int\delta(t_1+\ldots+t_n-t)\prod_{k=1}^n(3\exp(2t_k-e^{3t_k})dt_k)   $$
$$	=\int\cdots\int\delta(t_1+\ldots+t_n-t)\prod_{k=1}^n(3\exp(3t_k-e^{3t_k})dt_k)   $$
$$	=\int\cdots\int\delta(t_1+\ldots+t_n-t)\prod_{k=1}^n(\phi(t_k)dt_k)=\phi^{*n}(t)\eqno{(13)}   $$
hence
$$	e^tG_n(e^t)=e^t\int_t^\infty ds\,e^{-s}\phi^{*n}(s)=(\phi^{*(n-1)}*\psi)(t)\eqno{(14)}   $$
\par
	We discuss now more precisely the distribution $G_n(y)dy$ obtained above.  For $t\to-\infty$, $\phi(t)\approx3\eta_3(|t|)$ [where $\eta_m(t)=\theta(t)e^{-mt}$ as defined in Section 2], and $\psi(t)\approx\Gamma(2/3)e^{-|t|}$, so that
$$	\phi*\psi(t)\approx3\Gamma({2\over3})\int\theta(|t|-s)\theta(s)e^{-(|t|-s)}e^{-3s}\,ds
	=3\Gamma({2\over3})e^{-|t|}\int_0^{|t|}e^{-2s}\,ds   $$
$$	={3\over2}\Gamma({2\over3})e^{-|t|}(1-e^{-2|t|})\approx{3\over2}\Gamma({2\over3})e^{-|t|}   $$
By induction, $\phi^{*(n-1)}*\psi(t)\approx(3/2)^{n-1}\Gamma(2/3)e^{-|t|}$ when $t\to-\infty$.  Therefore $G_n(y)\to(3/2)^{n-1}\Gamma(2/3)$ when $y\to0$, i.e.,
$$	G_n(0)=(3/2)^{n-1}\Gamma(2/3)=(3/2)^n\Gamma(5/3)\eqno{(15)}   $$.  
\par
	If we write $\tilde\phi(t)=e^{-t}\phi(t)$, (13) gives
$$	-{d\over dt}G_n(e^t)
	=\int\cdots\int\delta(t_1+\ldots+t_n-t)\prod_{k=1}^n(\tilde\phi(t_k)dt_k)=\tilde\phi^{*n}(t)\eqno{(16)}   $$
so that
$$	-{d\over dy}G_n(y)={1\over y}\tilde\phi^{*n}(\ln y)\ge0   $$
i.e., $G_n(y)$ is a decreasing function of $y$.
\medskip
We study now the behavior of $G_n(y)$ for large $y$.  We have
$$	\tilde\phi^{*n}(t)=\int dt_1\cdots\int dt_n\,\delta(t_1+\ldots+t_n-t)\prod_{k=1}^n(3\exp{(2t_k-e^{3t_k})})   $$
$$	=e^{2t}\int dt_1\cdots\int dt_n\,\delta(t_1+\ldots+t_n-t)\prod_{k=1}^n(3\exp{(-e^{3t_k})})   $$
$$	=3e^{2t}\int d\tau_1\cdots\int d\tau_n\,\delta(\tau_1+\ldots+\tau_n-3t)\exp{(-\sum_{k=1}^ne^{\tau_k})}   $$
$$ =3e^{2t}\int d\tau'_1\cdots\int d\tau'_n\,\delta(\tau'_1+\ldots+\tau'_n)\exp{(-e^{3t/n}\sum_{k=1}^ne^{\tau'_k})} $$
A quadratic approximation of $\sum e^{\tau'_k}$ gives
$$	\sum_{k=1}^ne^{\tau'_k}\approx\sum_{k=1}^n(1+\tau'_k+{1\over2}{\tau'_k}^2)
	=n+{1\over2}\sum_{k=1}^n{\tau'_k}^2=n+{1\over2}[\sum_{k=1}^{n-1}{\tau'_k}^2+(\sum_{k=1}^{n-1}\tau'_k)^2]
	=n+{1\over2}\tau'^TA\tau'   $$
where $\tau'\in{\bf C}^{n-1}$ has components $\tau'_1,\ldots,\tau'_{n-1}$ and $A$ is a $(n-1)\times(n-1)$ matrix with $A_{k\ell}=2$ if $k=\ell$, $=1$ otherwise.  One finds $\det A=n$, and we have thus
$$	\tilde\phi^{*n}(t)\approx 3e^{2t}\int d\tau'_1\cdots\int d\tau'_{n-1}\exp[-e^{3t/n}(n+{1\over2}\tau'^TA\tau')]   $$
$$	=3e^{2t}\exp(-ne^{3t/n})\Big({(2\pi)^{n-1}\over ne^{3t(n-1)/n}}\Big)^{1/2}
	=3\Big({(2\pi)^{n-1}\over n}\Big)^{1/2}e^{t(3/n+1)/2}\exp(-ne^{3t/n})   $$
Therefore, by (16),
$$	-{dG_n\over dy}\approx3\Big({(2\pi)^{n-1}\over n}\Big)^{1/2}y^{(3/n-1)/2}\exp(-ny^{3/n})   $$
and finally for large $y$
$$	G_n(y)\approx\Big({(2\pi)^{n-1}\over n}\Big)^{1/2}(y^{(1-3/n)/2}+\ldots)\exp(-ny^{3/n})\eqno{(17)}   $$
$$	\int_Y^\infty G_n(y)\, dy
	\approx{1\over3}\Big({(2\pi)^{n-1}\over n}\Big)^{1/2}(Y^{3(1-3/n)/2}+\ldots)\exp(-nY^{3/n})   $$
In particular
$$	G_1(y)\approx y^{-1}\exp(-y^3)\qquad,\qquad G_2(y)\approx\sqrt\pi\,y^{-1/4}\exp(-2y^{3/2})   $$
$$	G_3(y)\approx{2\pi\over\sqrt3}\exp(-3y)\qquad,\qquad G_4(y)\approx{(2\pi)^{3/2}\over2}y^{1/8}\exp(-4y^{3/4})   $$
\medskip\noindent
{\bf 4. The distribution of small velocity increments.}
\medskip
	Instead of $F_\Delta(u)du$, which approximates the distribution of the $x$-component $u$ of $\Delta_r{\bf v}$, it is convenient to consider the distribution $p_n(z)dz$ of $z=u({\rm Var}(F_\Delta))^{-1/2}$ so that Var$(p_n)=1$.  We claim that in our case $y=((1/3)\Gamma(5/3)^n)^{1/2}|z|$, so that
$$	p_n(z)={1\over2}\Big({1\over3}\Gamma\Big({5\over3}\Big)^n\Big)^{1/2}
	G_n\Bigg(\Big({1\over3}\Gamma\Big({5\over3}\Big)^n\Big)^{1/2}|z|\Bigg)\eqno{(18)}   $$
\Big[Since Var$(p_n)=1$ by definition we have to estimate Var$(G_n)$.  We find, using formula (16),
$$	3{\rm Var}(G_n)=\int_0^\infty3y^2G_n(y)\,dy=-\int y^3{dG_n\over dy}\,dy=\int y^2\tilde\phi^{*n}(\ln y)\,dy   $$
$$=\int_{-\infty}^\infty\tilde\phi^{*n}(t)e^{3t}\,dt=\Big(\int\tilde\phi(t)e^{3t}\,dt\Big)^n=\Big(\int\phi(t)e^{2t}\,dt\Big)^n$$
$$	=\Big(\int3e^{5t}\exp(-e^{3t})\,dt\Big)^n=\Big(\int_0^\infty w^{2/3}e^{-w}\,dw\Big)^n
	=\Gamma\Big({5\over3}\Big)^n  $$
where $\Gamma(5/3)\approx0.9027452929$.\Big]
\medskip
	A natural guess is that the distribution $G_n(y)dy$ corresponding to the distribution $p(z)dz$ measured at Rey\-nolds number ${\cal R}$  is obtained when $\ell_n(=\kappa^{-n}\ell_0)$ is equal to the Kolmogorov dissipation length $\approx\ell_0{\cal R}^{-3/4}$ so that in our model ${\cal R}^{-3/4}\approx\ell_n/\ell_0=\kappa^{-n}$ or $\kappa^n={\cal R}^{3/4}$ or $n\ln\kappa=(3/4)\ln {\cal R}$, or
$$	n={3\over4}{\ln{\cal R}\over\ln\kappa}\approx0.24\ln {\cal R}\eqno{(19)}   $$
with the estimate $(\ln\kappa)^{-1}=0.32$ in [12].  This gives $n=1.10$ for ${\cal R}=96$ and $n=2.03$ for ${\cal R}=4638$.
\medskip
	Since $G_n(0)=(3/2)^n\Gamma(5/3)$ by (15), we obtain from (18)
$$	p_n(0)={1\over2}\Big({1\over3}\Gamma\Big({5\over3}\Big)^n\Big)^{1/2}\Big({3\over2}\Big)^n
	\Gamma\Big({5\over3}\Big)={1\over2\sqrt3}\Gamma\Big({5\over3}\Big)\cdot
	\Big({3\over2}\Gamma\Big({5\over3}\Big)^{1/2}\Big)^n   $$
Here we have
$$	{1\over2\sqrt3}\Gamma\Big({5\over3}\Big)=0.260600118\qquad{\rm and}\qquad
	{3\over2}\Gamma\Big({5\over3}\Big)^{1/2}=1.425193638   $$
The above prediction for $p(0)$ can be compared with experimental data (see [13]).
\par
	For small $y>0$ we have
$$	-G'_n(y)={1\over y}\tilde\phi^{*n}(\ln y)\approx{1\over y}3^n\eta_2^{*n}(|\ln y|)={1\over y}3^ne^{-2|\ln y|}{|\ln y|^{n-1}\over(n-1)!}={3^n\over(n-1)!}y|\ln y|^{n-1}   $$
from which one can estimate
$$	p'_n(z)={1\over6}\Gamma\Big({5\over3}\Big)^n\,
	G_n'\Bigg(\Big({1\over3}\Gamma\Big({5\over3}\Big)^n\Big)^{1/2}|z|\Bigg)   $$
for small $z>0$.
\medskip\noindent
{\bf 5. The distribution of large velocity increments.}
\medskip
	We consider now the problem of comparing the distributions $p_n(z)dz$ with distributions obtained in (numerical) experiments [13] for large $z$.  Note that $\ln p_n(z)$ corresponds via (18) to $\ln G_n(y)\approx-ny^{3/n}$ (see (17)).  Superficially, the curves in [13] (for instance Fig. 6) look like $\ln p_2$, $\ln p_3$, $\ln p_4$, passing from a concave ($\ln p_2$) to a linear ($\ln p_3$) then a convex ($\ln p_4$) behavior at large $z$ when ${\cal R}$ increases.  Things are however not quite that simple.
\medskip
	The relation (19) gives $n=2.03$ for ${\cal R}=4638$ and this does not compare well with the data of [].  In fact $n=2$ gives a concave function $\log p_2$ while for ${\cal R}=4638$ Fig. 6 of [] gives $\log p$ linear or convex at large $z$.  To resolve this conflict we shall be more careful and take into account the fluctuations of $n$ at given ${\cal R}$.
\medskip
	In the presence of viscosity the absolute value of the radial velocity gradient is
$$	\omega=|{\partial v_x\over\partial x}|\approx{\Delta_r v\over r}\hbox
	{ when $r$ is the dissipation length}   $$
We fix $\ell_0,V_0$, write $\omega_0=V_0^{1/3}/\ell_0$, and define $P(\xi)\,d\xi$ to be the probability distribution of $\xi=\omega/\omega_0$.
\medskip
	As in Section 2 let $\varpi$ denote the probability measure on sequences $(V_n)_1^\infty$ defined by (3).  Then with high $\varpi$-probability the sequence $(V_1,\ldots,V_n,\ldots)$ decreases to $0$.  Writing again $S_j=\{(V_n):V_j^{1/3}\ell_j>\nu\}$ we have thus with high probability $S_1\supset\ldots\supset S_n\supset\ldots$ and (6) shows that $\varpi(S_n)$ decreases exponentially with $n$.
\medskip
	We use the near partition into the sets $S_{n-1}\backslash S_n$ to approximate $P(\xi)$ as
$$	P(\xi)\approx\sum_nP_n^*(\xi)   $$
with
$$	P_n^*(\xi)=\ell_n\omega_0\Big(\prod_{k=1}^n\int{\kappa\,dV_k\over V_{k-1}}e^{-V_k/V_{k-1}}\Big)^*{1\over V_n^{1/3}}\chi_{[0,V_n^{1/3}]}(\ell_n\omega_0\xi)   $$
where $(\ldots)^*$ means that the integrals are restricted to $S_{n-1}\backslash S_n$.  [Note that there is some arbitrariness in computing the gradient $\xi$ at level $n$ for a viscosity cutoff between level $n-1$ and level $n$].
\medskip
	Suppose now that for some $j$ we have
$$	{\nu\over\ell_j}\le\ell_j\omega_0\xi\qquad{\rm i.e.}\qquad\kappa^{2j}\le\xi{\cal R}   $$
Then if $n\le j$ we find $P_n^*(\xi)=0$ because
$$	\Big({\nu\over\ell_n}\le\ell_n\omega_0\xi\qquad{\rm and}\qquad V_n^{1/3}\le{\nu\over\ell_n}\Big)
	\qquad\Rightarrow\qquad\chi_{[0,V_n^{1/3})}(\ell_n\omega_0\xi)=0   $$
We have thus
$$	P(\xi)\approx\sum_{n>j}P_n^*(\xi)\qquad{\rm if}\qquad\kappa^{2j}\le\xi{\cal R}   $$
Note that if we remove the $*$ in the definition of $P_n^*$ we have
$$	P_n(\xi)=\ell_n\omega_0.2F_\Delta(\ell_n\omega_0\xi)
	=\ell_n\omega_0(\kappa^n/V_0)^{1/3}G_n((\kappa^n/V_0)^{1/3}\ell_n\omega_0\xi)   $$
$$	=\kappa^{-2n/3}G_n(\kappa^{-2n/3}\xi)
	\approx\Big({(2\pi)^{n-1}\over n}\Big)^{1/2}\kappa^{1-n}\xi^{(1-3/n)/2}\exp(-n(\kappa^{-2n/3}\xi)^{3/n})   $$
which shows that the variable $\xi$ in $P$ corresponds to $y=\kappa^{-2n/3}\xi$ in $G_n$ so that the values of $\xi$ corresponding to the abscissa $z$ in [13] are quite large.
\medskip
	We consider in particular the case ${\cal R}=4638$ in Fig. 6 of [13].  The condition $\kappa^{2j}\le\xi{\cal R}$ becomes $\xi\ge34.5$ for $j=2$ (with $\kappa=20$) so that, except for $z$ close to $0$ only the $P_n^*$ with $n>2$ contribute.  This gives a reasonable interpretation of the fact that $\log_{10}p(z)$ appears to be dominated by $P_3$, with some admixture of $P_4$ visible at the largest $z$.
\medskip
	The above discussion is somewhat qualitative, based on an analytic study of the measure $\varpi$ defined by (3).  Numerical studies, using (3) and the cutoff $V_n^{1/3}\ell_n>\nu$ are desirable, to test quantitatively how accurate a description of turbulence by the measure $\varpi$ really is.
\vfill\eject
\medskip\noindent
{\bf 6. Comparison with the ideas of Kolmogorov and Obukhov.}
\medskip
	We note that, with $\alpha(t)=\exp(t-e^t)$ as in Section 2, we have $\phi(t)=3\alpha(3t)$ so that
$$	\int t\phi(t)\,dt=-{\gamma\over3}\qquad,\qquad\int t^2\phi(t)\,dt={1\over9}(\gamma^2+{\pi^2\over6})   $$
and also
$$	\int t\psi(t)\,dt=\int te^t\,dt\int_t^\infty e^{-s}\phi(s)\,ds=\int dt\,[e^t(t-1)]e^{-t}\phi(t)
	=\int dt\,(t-1)\phi(t)=-{\gamma\over3}-1   $$
$$	\int t^2\psi(t)\,dt=\int t^2e^t\,dt\int_t^\infty e^{-s}\phi(s)\,ds=\int dt\,[e^t(t^2-2t+2)]e^{-t}\phi(t)
	=\int dt\,(t^2-2t+2)\phi(t)   $$
$$	=({\gamma\over3}+1)^2+{1\over9}\cdot{\pi^2\over6}+3   $$
Therefore ${\rm Var}(\phi)=\pi^2/54$ and ${\rm Var}(\psi)=\pi^2/54+3$.
\medskip
	The central limit theorem applied to (14) implies then that $e^tG_n(e^t)$ is asymptotically Gaussian for large $n$:
$$	e^tG_n(e^t)\approx{1\over\sigma\sqrt{2\pi}}\exp\big(-{(t+\bar\mu)^2\over2\sigma^2}\big)   $$
with $\bar\mu=n\gamma/3+1$, $\sigma^2=n\pi^2/54+3$.  This means that $G_n(y)\,dy$ is asymptotically a lognormal distribution for large $n$.  Note however that $\alpha(t)$, $\phi(t)$, $\psi(t)$ have only exponential (not Gaussian) decay when $t\to-\infty$.  The deviation of the $G_n$ from lognormal explains why the exponents $\tau_p$, $\zeta_p$ are not correctly predicted by lognormal theory.
\medskip
	The necessity to take fluctuations into account in Kolmogorov's classical theory of turbulence (Kolmogorov [7], [8], [9]) was pointed out by Landau.  In [10], Kolmogorov used ideas of Obukhov to deal with Landau's remark, and presented a lognormal theory of fluctuations of the energy dissipation $\epsilon_\ell$ on scale $\ell$.
\medskip
	In our approach the probability Ansatz (3) is fundamental (based on a statistical mechanical understanding of turbulence) and the lognormality of $\epsilon_\ell$ is an approximate deduction.  This explains why the exponents $\tau_p$, $\zeta_p$ obtained from (3) are different from those obtained from a lognormal theory (and give a better fit of the experimental data).
\medskip
	Christian Beck, who has studied a hierarchical model of turbulence [2], has conjectured that in such a hierarchical model one could derive a lognormal distribution for $\Delta_rv$ [private communication in December 2013 during a workshop at the Isaac Newton Institute in Cambridge, UK].  The present paper provides an example of such a derivation.
\vfill\eject
\noindent
{\bf References.}
\medskip
[1] F. Anselmet, Y. Gagne, E.J. Hopfinger, and R.A. Antonia ``High-order
velocity structure functions in turbulent shear flows.''  J. Fluid
Mech. {\bf 140},63-89(1984).

[2] C. Beck.  ``Chaotic cascade model for turbulent velocity distributions.''  {\it Phys. Rev. E} {\bf 49},3641-3652(1994).

[3] R. Benzi, G. Paladin, G. Parisi, and A. Vulpiani ``On the multifractal
nature of fully developed turbulence and chaotic systems.''J. Phys. A {\bf
17},3521-3531(1984).

[4] B. Castaing.  ``The temperature of turbulent flows.''  {\it J. Phys. II France} {\bf 6},105-114 (1996).

[5] U. Frisch and G. Parisi ``On the singularity structure of fully developed turbulence'' in {\it Turbulence and Predictability in Geophysical Fluid Dynamics} (ed. M. Ghil, R. Benzi, and G. Parisi), pp. 84-88.
North-Holland, 1985.

[6] G. Gallavotti {\it Foundations of Fluid Mechanics.}  Springer-Verlag,
Berlin, 2005 (see Section 6.3).

[7] A.N. Kolmogorov.  ``The local structure of turbulence in incompressible viscous fluid for very large Reynolds number.''  Dokl. Akad. Nauk SSSR {\bf 30},301-305(1941).

[8] A.N. Kolmogorov.  ``On degeneration (decay) of isotropic turbulence in an incompressible viscous liquid.''  Dokl. Akad. Nauk SSSR {\bf 31},538-540(1941).

[9] A.N. Kolmogorov.  ``Dissipation of energy in locally isotropic turbulence.''  Dokl. Akad. Nauk SSSR {\bf 32},16-18(1941).

[10] A.N. Kolmogorov.  ``A refinement of previous hypotheses concerning the local structure of turbulence in a viscous incompressible fluid at high Reynolds number.''  {\it J. Fluid Mech.} {\bf 13},82-85(1962).

[11] C. Meneveau and K.R. Sreenivasan ``Simple multifractal cascade model
for fully developed turbulence.''  Phys. Rev. Lett. {\bf 59},1424-1427(1987).

[12] D. Ruelle.  ``Hydrodynamic turbulence as a problem in nonequilibrium statistical mechanics.''  PNAS {\bf 109},20344-20346(2012).

[13] J. Schumacher, J. Scheel, D. Krasnov, D. Donzis, K. Sreenivasan, and V. Yakhot.  ``Small-scale universality in turbulence.''  Preprint (2014).

[14] R. Stresing and J. Peinke.  ``Towards a stochastic multi-point description of turbulence.''  {\it New J. of Phys.} {\bf 12},103046+14(2010).

[15] V. Yakhot ``Pressure-velocity correlations and scaling exponents in
turbulence.''  J. Fluid Mech. {\bf 495}135-143(2003).
\end